\documentclass[aps,prb,reprint,twocolumn,showpacs,floatfix,superscriptaddress]{revtex4-2}

\usepackage{amssymb,amsmath,amstext}                
\usepackage{graphicx}                                               
\usepackage{epstopdf}                                               
\usepackage{color}                                                     
\usepackage{bm}                                                        
\usepackage{appendix}                                              
\usepackage[utf8]{inputenc}
\usepackage{latexsym}
\usepackage{xcolor}
\usepackage{braket}
\usepackage{ulem}
\usepackage{siunitx}
\usepackage{umoline}
\usepackage{epsfig}
\usepackage{graphics}
\usepackage{amssymb}
\usepackage{psfrag}
\usepackage{mathtools}
\usepackage[english]{babel}
\usepackage[T1]{fontenc}
\usepackage{lmodern}
\usepackage{booktabs}

\usepackage{rotating}
\usepackage{multirow}

\definecolor{lblue} {RGB}{51,71,158}
\usepackage[colorlinks=true,citecolor=blue,linkcolor=blue,urlcolor=lblue]{hyperref}

\begin{document}

\title{ Slow dynamics of a mobile impurity interacting with an Anderson insulator}

\author{Piotr Sierant} 
\email{piotr.sierant@icfo.eu}
\affiliation{ICFO-Institut de Ci\`encies Fot\`oniques, The Barcelona Institute of Science and Technology, Av. Carl Friedrich
Gauss 3, 08860 Castelldefels (Barcelona), Spain}

\author{Titas Chanda}
\email{tchanda@ictp.it}
\affiliation{The Abdus Salam International Centre for Theoretical Physics (ICTP), Strada Costiera 11, 34151, Trieste, Italy}

\author{Maciej Lewenstein} 
\email{maciej.lewenstein@icfo.eu}
\affiliation{ICFO-Institut de Ci\`encies Fot\`oniques, The Barcelona Institute of Science and Technology, Av. Carl Friedrich
Gauss 3, 08860 Castelldefels (Barcelona), Spain}
\affiliation{ICREA, Passeig Lluis Companys 23, 08010 Barcelona, Spain}

\author{Jakub Zakrzewski} 
\email{jakub.zakrzewski@uj.edu.pl}
\affiliation{Instytut Fizyki Teoretycznej, 
Uniwersytet Jagiello\'nski,  \L{}ojasiewicza 11, PL-30-348 Krak\'ow, Poland}
\affiliation{Mark Kac Complex Systems Research Center, Uniwersytet Jagiello{\'n}ski, Krak{\'o}w, Poland}

\date{\today}

\begin{abstract}
We investigate dynamics of a single mobile impurity immersed in a bath of Anderson localized particles and focus on the regime of relatively strong disorder and interactions. In that regime, the dynamics of the system is particularly slow, suggesting, at short times, an occurrence of  many-body localization. Considering longer time scales, we show that the latter is a transient effect and that, eventually, the impurity spreads sub-diffusively and induces a gradual delocalization of the Anderson insulator. The phenomenology of the system in the considered  regime of slow dynamics includes a sub-diffusive growth of mean square displacement of the impurity, power-law decay of density correlation functions of the Anderson insulator and a power-law growth of entanglement entropy in the system. We observe a similar regime of slow dynamics also when the disorder in the system is replaced by a sufficiently strong quasi-periodic potential. 
\end{abstract}

\maketitle

\section{Introduction}
Dynamics of quantum many-body systems is actively investigated in synthetic quantum matter such as ultracold atoms \cite{Bloch08, Lewenstein07}.
Eigenstate thermalization hypothesis (ETH) \cite{Deutsch91,Srednicki94,Rigol08, Alessio16} describes how quantum systems, which are initially prepared in far-from-equilibrium states, evolve in time to an equilibrium state. 
The equilibrium state is determined by the global constants of motion such as the the energy, total momentum etc., but otherwise, it is independent of the details of the initial state. This ergodic behavior of quantum many-body systems is similar, in spirit, to thermalization of isolated classical systems that explore all possible configurations allowed by global conservation laws.
Importantly, the investigations of dynamics of quantum many-body systems have provided us with counterexamples to ETH, i.e. with classes of systems that do not thermalize under their own dynamics, such as: integrable models \cite{Vidmar16}; strongly disordered, many-body localized (MBL) systems \cite{Gornyi05,Basko06, Pal10, Ros15, Imbrie16, Alet18, Abanin19}; disorder-free systems such as tilted lattices \cite{Schulz19, vanNieuwenburg19, Taylor20, Chanda20c, Yao20b, Scherg21, Guo21, Morong21, Yao21, Yao21a}; models with global constraints \cite{Turner18,Feldmeier19,Nandkishore19,Sala20,Khemani20,Szoldra22}; or implementations of lattice gauge theories \cite{Robinson19,James19,Surace19,Chanda20,Banuls20,Aidelsburger22,Aramthottil22}.

Quantum impurity models, which describe a small quantum system coupled to a reservoir of particles, appear naturally in various situations, constituting another class of systems whose dynamics is a fundamental problem in many-body physics, see e.g. \cite{Leggett87, Hewson93}. One particular example of a quantum impurity system is an electron
in a dielectric crystal. The motion of the electroc distorts the spatial configuration of its surrounding, which effectively screens its charge.
The resulting system, which consists of the electron and the surrounding phonon cloud, is called a polaron, as coined by Pekar \cite{Pekar46, Pekar47,Pekar48}, see also \cite{Landau33, Landau48}. The concept of polaron has been extended to describe a generic particle, the impurity, in a generic material, e.g. a conductor, a semiconductor or a gas \cite{Froehlich54,Alexandrov}. 

One important example is that of an impurity embedded
in an ultracold gas. This system has been widely studied
both theoretically and experimentally, in the case
of a ultracold Fermi \cite{Schirotzek09, Nascimbene09, Sommer11, Kohstall12,Frohlich11, Koschorreck12,Massignan14,Yan19} or Bose gas \cite{Cote02,Massignan05,Cucchietti06,Palzer09,Catani12,Spethmann12,Bonart12,Rath13, Fukuhara13q, Shashi14,Meinert17}. The problem of the impurity in a Bose-Einstein condensate \cite{Hu16a, Jorgensen16, Zoe20,Mathy12, Schecter12,Massel13, Charalambous19,Charalambous20} can be viewed as an example of Quantum Brownian motion \cite{Lampo17, Mehboudi19,Khan21,Khan22}.
Recently, there has been a considerable interests in studying impurities in strongly correlated systems with topological order, such as fractional quantum Hall systems \cite{Grass20} or topological Mott insulators 
\cite{Farre20}.

The reach physics of quantum impurity models, in connection with the phenomenon of MBL, gives rise to new dynamical phenomena. Typically, coupling to external baths delocalizes the system \cite{Nandkishore17a}, as shown for bosonic baths \cite{Bonca18} or for $t-J$ model \cite{Bonca17,Gopalakrishnan17,Lemut17}, leading to a subdiffusive \cite{Prelovsek18} (or diffusive) spread of the initially localized particle. Interestingly, the localization in the system can be restored by by driving the system to a non-equilibrium steady state in presence of a coupling to a bath \cite{Lenarcic18}.

Recently, a scenario of a single mobile impurity interacting with a system of Anderson localized particles was considered in a series of works \cite{Krause21, Brighi21a,Brighi21b}. It was shown that, in certain parameter regimes, the impurity acts as an ergodic seed enabling thermalization of the system. In contrast, for sufficiently large disorder strength $W$ and sufficiently strong coupling $U$ between the impurity and the Anderson insulator, the mobile impurity becomes localized \cite{Brighi21b} and the system becomes MBL \cite{Brighi21a}. The investigations of time evolution in \cite{Brighi21a,Brighi21b} have been performed up to time $t_1=200 J^{-1}$ where $J$ denotes the nearest neighbor tunneling amplitude. While the time scale $t_1$ is significant from the perspective of experiments with ultracold atoms in optical lattices (note that times of the order of $1000 J^{-1}$ are presently reachable \cite{Scherg21}), it appears to be relatively short given the strong interactions $U$ and large disorder strength $W$ in the system. Moreover, the recent study \cite{Sierant22} of time evolution in disordered XXZ spin chain shows the crucial role of taking the limit of long times when assessing whether the system is MBL. This limit is directly connected to strong finite size effects at the MBL transition, which prevent present exact numerical studies from an unambiguous answer to a question of whether a stable dynamical MBL phase can be observed in the thermodynamic limit \cite{Suntajs20e, Sierant20b,Abanin21, Panda20, Kiefer20, Sels20, Sels21a, Kiefer21, Sierant21,Ghosh22, Sierant22f} and what is the critical disorder strength for the transition \cite{Sierant20p, Morningstar22, Sels21}.

Motiviated by the importance of quantum impurity problems, their experimental relevance
\cite{Rubio18}, and by the recent controversies around MBL, we revisit the problem of mobile impurity interacting with an Anderson insulator \cite{Krause21, Brighi21a,Brighi21b}.
We show that extending the analysis of time evolution to longer times yields a different perspective on the localization in the system. We find, that the impurity boson, after an initial transient dynamics suggesting its localization, eventually spreads sub-diffusively
across the system inducing its slow thermalization and possibly shifting the boundary of the MBL regime to larger $W$ and $U$ than anticipated in \cite{Brighi21a, Brighi21b}.

The remainder of this work is structured as follows. In Sec.~\ref{sec:model} we give a precise definition of the considered model of mobile impurity interacting with Anderson insulator. In Sec.~\ref{sec:imp} we present our numerical results showing the spreading of the impurity in the system. Subsequently, in Sec.~\ref{sec:cor} we show the impact of the spreading impurity on the Anderson insulator and discuss the growth of entanglement entropy in the system in Sec.~\ref{sec:ent}. Slow dynamics of the system in presence of quasiperiodic potential is investigated in Sec.~\ref{sec:qpd}.  We conclude in Sec.~\ref{sec:con} and show details about convergence of our numerical calculations in App.~\ref{app:1}.

\section{The model}
\label{sec:model}

We consider the model of mobile impurity interacting with Anderson insulator  described by the Hamiltonian
\begin{eqnarray}
 H&=& J\sum_{i=1}^{L-1} (\hat{d}_i^\dagger \hat{d}_{i+1} + H.c.) + \sum_{i=1}^L h_i \hat{n}_{d,i}  \cr
 &+& J\sum_{i=1}^{L-1} (\hat{c}_i^\dagger \hat{c}_{i+1} + H.c.) +U   \sum_{i=1}^L  \hat{n}_{c,i}\hat{n}_{d,i},
 \label{eq:hamb}
\end{eqnarray}
where $\hat{d}_i, \hat{d}_i^\dagger$ and $\hat{c}_i, \hat{c}_i^\dagger$ are annihilation and creation operators respectively of clean ($c$) or disorder ($d$) hardcore boson at site $i$, whereas $\hat{n}_{x,i}$ with  $x \equiv c,d$ are the corresponding occupation number operators. 
The tunneling amplitude $J \equiv 1$ is fixed as the energy unit, 
$U$ denotes the interaction strength between the two species of bosons, and the on-site potential $h_i$ are independent random variables taken from uniform distribution on the interval $[-W,W]$ where $W$ is the disorder strength, and $L$ is the system size.

In absence of interactions ($U=0$), the $d$-bosons form an Anderson insulator, whereas the $c$-boson propagates freely on the lattice. When the interactions are present ($U \neq 0$), the system is in one of a two distinct dynamical regimes. For relatively weak disorder and interactions, e.g. $W=1$ and $U=1$, the system thermalizes due to to correlated hops of the $c$ bosons and the localized $d$ bosons \cite{Krause21}. In that regime, the impurity becomes an ergodic seed which propagates throughout the lattice delocalizing the system. In contrast, for stronger disorder and interactions,
the $c$-boson may remain exponentially localized around its initial position \cite{Brighi21b}. This, in consequence, induces a logarithmic growth of entanglement entropy \cite{Brighi21a} reminiscent of the behavior of MBL systems \cite{Znidaric08, Bardarson12, Serbyn13a,iemini2016signatures}.

In this work we focus on the slow dynamics of the system in the latter regime. We consider two initial states: i) $\ket{\psi_{cent}(0)}$ in which  the $c$-boson is placed initially in the middle of the chain (at $i_0=L/2$) and the $d$-bosons form a density wave state $\ket{0,1,0,0,1,0,0,1,..}$ at $1/3$ filling; ii) $\ket{\psi_{left}(0)}$ in which the $c$-boson is placed initially on the leftmost site of the chain (at $i_0=0$) and the $d$-bosons form a density wave state $\ket{0,0,1,0,0,1,0,0,1,..}$. The first initial configuration matches precisely the initial state considered in \cite{Brighi21a}, whereas the second allows for a longer distance of propagation of the $c$-boson, similarly to \cite{BarLev15}, enabling further insights into the dynamics of the system. We consider quantities averaged over $50$ disorder realizations for $L=60$ (similar number was considered in \cite{Brighi21a}) and over $200$ disorder realizations for $L\leq 30$.

We also investigate a situation when the disorder \eqref{eq:hamb} is introduced via quasiperiodic potential, for which $h_j = W_{QP} \cos(2\pi k j + \phi)$, where $k=(\sqrt{5}-1)/2$ and $\phi$ is a random phase taken from the uniform distribution between  $[0,2\pi]$.
In that case, in the absence of interactions ($U=0$), the system of $d$-bosons undergoes Aubry–André localization  at $W_{QP} > 2$ \cite{Aubry80}. In presence of density-density interactions between the $d$-bosons, this system  behaves similarly to disordered Heisenberg spin chain 
\cite{Iyer13, Naldesi16, Setiawan17,BarLev17, Bera17a,Weidinger18,Doggen19, Weiner19,Singh21,Aramthottil21}, although certain features, such as an appearance of persistent oscillations of the imbalance  at large $W_{QP}$ \cite{Sierant22}, are specific to the quasiperiodic potential. In this work, we focus on the regime of slow dynamics that occurs at $W_{QP}=6$  and interaction between the $c$-bosons and $d$-boson taken as $U=12$.

To calculate the time evolution generated by the Hamiltonian \eqref{eq:hamb}, we use two complementary approaches. For smaller system sizes, we employ expansion of the evolution operator $e^{-i H t}$ into Chebyshev polynomials \cite{Fehske08}. This approach relies on the sparse structure of the Hamiltonian matrix \eqref{eq:hamb} and allows us to obtain numerically exact results for system comprised of up to $L=30$ sites, i.e. with dimension of the Hilbert space $\approx 901 \cdot 10^6$. For larger system sizes ($L=60$), we use tensor network based algorithms TEBD \cite{Vidal03} and TDVP \cite{Haegeman11, Koffel12, Haegeman16, Paeckel19,Goto19} implemented in ITensor \cite{itensor2, itensor2a}. In the latter case, we check the convergence of our results with respect to the  bond dimension $\chi$, details are given in App.~\ref{app:1}.

 \begin{figure*}
    \includegraphics[width=0.99\linewidth]{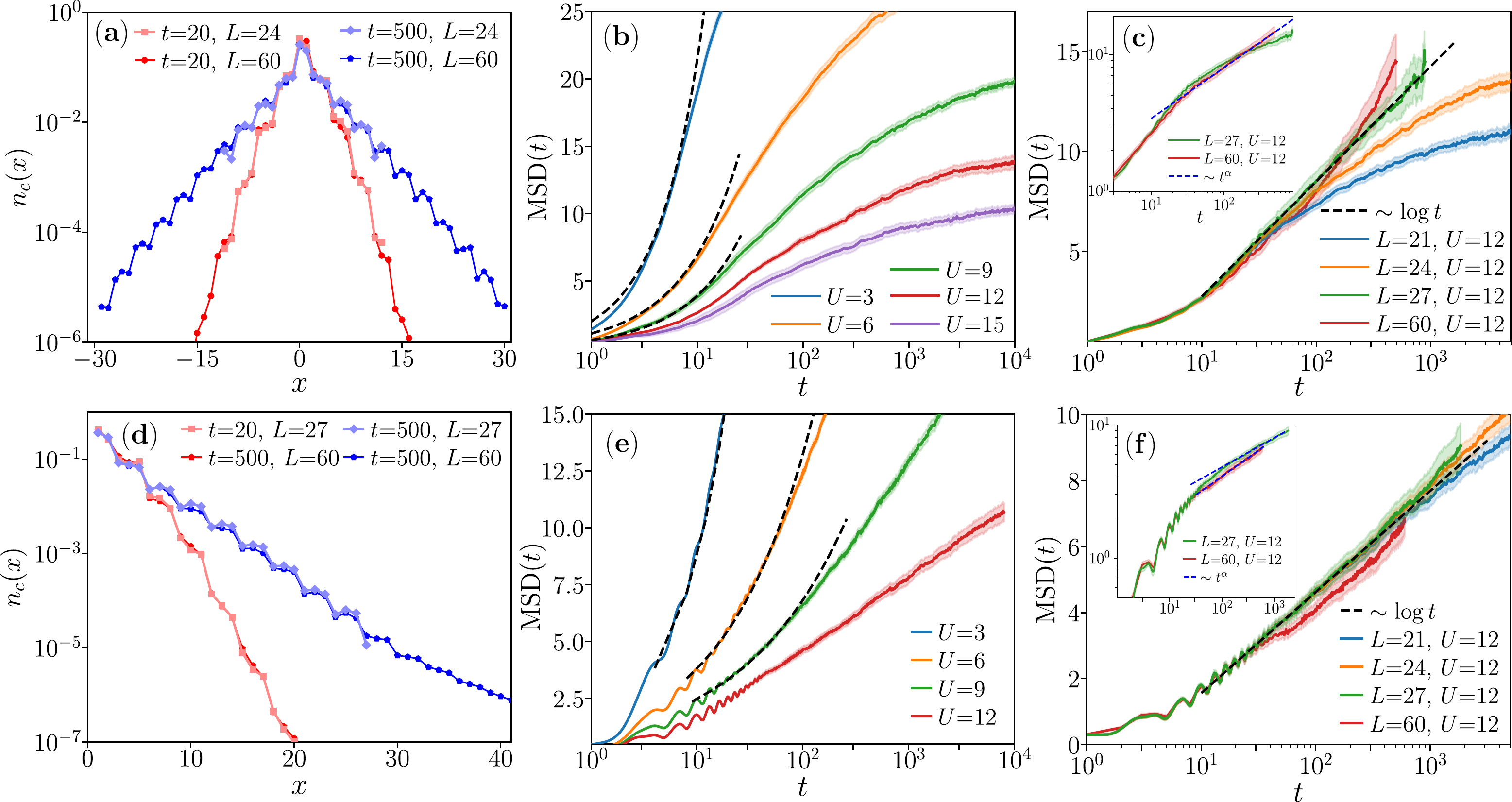} \vspace{-0.1cm}
  \caption{Slow dynamics of the $c$-boson. The top row show the results for the $c$-boson starting in the center site of the chain; (\textbf{a}): average  density  $\braket{\hat{n}_{i,c}}$ of the $c$-boson
  for interaction strength $U=12$, disorder strength $W=6.5$, times $t=20, 500$ and system sizes $L=24, 60$, where $x$ is the number of the site with respect to the initial position of the $c$-boson, $x \equiv i - i_0$; (\textbf{b}): the MSD of the $c$-boson for system size $L=24$ as function of time $t$ for various interaction amplitudes $U$ at fixed disorder strength $W=6.5$, the black dashed lines show the power-law fits $MSD(t) \sim t^{\alpha}$ with $\alpha < 1$; (\textbf{c}): MSD for different system sizes $L$ at fixed $U=12$, $W=6.5$. The dashed line shows a logarithmic fit $MSD(t) \sim \log(t)$, whereas the inset shows data on log-log scale demonstrating that a power law fit $MSD(t) \sim t^{\alpha}$ works well for $L=60$ result. Panels (\textbf{d}), (\textbf{e}), (\textbf{f}): the same as (\textbf{a})-(\textbf{c}) but for the $c$-boson initially occupying the leftmost site of the chain. The shaded regions denote statistical uncertainties in the data associated with the finite number of disorder realizations. Data for $L=60$ was obtained with TDVP with bond dimension $\chi=800$.
  }\label{fig:clean_evo}
\end{figure*} 

\section{Dynamics of the impurity}
\label{sec:imp}

In this section we characterize the dynamics of the impurity, i.e. the $c$-boson. We start by investigating the disorder averaged density profile $\braket{\hat{n}_{i,c}}\equiv n_c(x)$ of the $c$-boson after certain time $t$ of time evolution, where $x\equiv i-i_0$ determines the position in the 1D lattice. The results for interaction amplitude $U=12$ and disorder strength $W=6.5$ are presented in Fig.~\ref{fig:clean_evo}(a) for the initial state $\ket{\psi_{cent}(0)}$ and Fig.~\ref{fig:clean_evo}(d) for $\ket{\psi_{left}(0)}$. In both cases we observe that the density profile of the $c$-boson is approximately exponentially decaying $\braket{\hat{n}_{i,c}}  \sim e^{-|x|/\xi(t)}$ where  $\xi(t)$ is time-dependent localization length, consistently with the results of \cite{Brighi21b}. On top of the overall exponential envelope, there are also small characteristic oscillations with period of 3 sites, associated with the initial density wave state of the $d$-bosons. 

In order to understand the dynamics of the impurity, we could try to investigate the time dependence of the localization length $\xi(t)$. However, in the course of slow delocalization of the system, the density profile $\braket{\hat{n}_{i,c}}$ gets increasingly non-exponential, see e.g. data for $L=60$ and $t=500$ in Fig.~\ref{fig:clean_evo}(d). For that reason, to have a more quantitative measure of the spread of the $c$-boson, we calculate the mean square dispacement (MSD) defined as
\begin{equation}
\mathrm{MSD}(t)\equiv \sum_{i=1}^L  (i- \overline i)^2 \braket{ \hat{n}_{i,c}}  ,
\label{dis}
\end{equation}
where $\overline i = \sum_{i=1}^L i \braket{  \, \hat{n}_{i,c} } $ is the mean position of the $c$-boson. The MSD is closely related to the second moment of density propagator studied in \cite{Bera17,Weiner19} in the context of MBL in disordered Heisenberg spin chain.  
The growth in time of $\mathrm{MSD}\sim t^\alpha$ would indicate a diffusive spreading of the $c$-boson for $\alpha=1$ and a sub-diffusive dynamics for $0<\alpha<1$. In contrast, for Anderson localized system, the density profile admits an exponentially decaying envelope with $\xi(t)$ saturating in time to localization length in the system, and, consequently MSD saturates in time to a disorder strength dependent constant.

We fix the disorder strength as $W=6.5$ and investigate time evolution of MSD varying the interaction strength $U$. The results for system size $L=24$ initialized in the states $\ket{\psi_{cent}(0)}$ and $\ket{\psi_{left}(0)}$ are shown in Fig.~\ref{fig:clean_evo}(b) and Fig.~\ref{fig:clean_evo}(d), respectively. For interaction amplitudes $U=3,6,9$, we observe an algebraic increase $\mathrm{MSD}(t) \sim t^{\alpha}$ where $\alpha \approx 1, 0.8, 0.8$ for $\ket{\psi_{cent}(0)}$ (and $\alpha \approx 0.9, 0.55, 0.45$ for $\ket{\psi_{left}(0)}$), showing that the $c$-boson spreads sub-diffusively throughout the lattice. We note that the powers $\alpha$ are smaller if the $c$-boson is initially placed at $i_0=0$ than in the center of the lattice $i_0=L/2$. Consequently, the interval of times in which the power-law increase of $\mathrm{MSD}(t)$ persists is wider for $\ket{\psi_{left}(0)}$. Moreover, when the system size is increased between $L=18$ and $L=24$, the subdiffusive growth of MSD persists to longer times and with exponent $\alpha$ that stays approximately constant or increases with $L$ (data not shown). Those results indicate that for $U \leq 9$ the $c$-boson remains mobile and propagates throughout the lattice, enabling the thermalization of the system \cite{Krause21}. 

The situation is not so clear-cut for $U=12$, the interaction strength considered in \cite{Brighi21a}, for which, in Fig.~\ref{fig:clean_evo}(b), we observe that MSD is saturating at a value significantly smaller than the maximal value allowed at $L=24$. At the same time, for $\ket{\psi_{left}(0)}$, in Fig.~\ref{fig:clean_evo}(d), we see an approximately logarithmic growth of MSD in time. The understanding of the dynamics in this regime of parameters is particularly challenging and requires a careful examination of the influence of system size on time evolution at late times. To that end, we show the time dependence of MSD$(t)$ for system sizes ranging from $L=21$ to $L=60$ in Fig.~\ref{fig:clean_evo}(c) and Fig.~\ref{fig:clean_evo}(f), for the initial states $\ket{\psi_{cent}(0)}$ and $\ket{\psi_{left}(0)}$, respectively.

When the $c$-boson is initially placed in the center of the lattice ($i_0 = L/2$), Fig.~\ref{fig:clean_evo}(c), we observe that, for $L=21, 24$, the mean square displacement grows logarithmically in time  MSD$(t) \sim \log(t)$ only in a narrow time interval $t\in[10,50]$ and then deviates from that behavior, which could suggest its localization. However, the time at which the deviation occurs increases with system size $L$, and already at for $L=27$, we observe that MSD$(t) \sim \log(t)$ in a much wider time interval $t\in[10,1000]$. Finally, for $L=60$, we observe that the MSD$(t)$ increases more quickly than logarithmically in time and is well fitted by a power-law MSD$(t) \sim t^{\alpha}$ with $\alpha \approx 0.37$. We note while the data for $L=21,24,27$ overlap at small times, this is not the case for $L=60$ due to a smaller number (50) of disorder realization used.

If the $c$-boson is initially placed at the leftmost site of the lattice ($i_0 = L/2$), Fig.~\ref{fig:clean_evo}(f), we observe that the logarithmic growth of MSD persist in the interval of times $t\in[10,2000]$ already for $L=21, 24$. The data for $L=27$ deviate from the logarithmic behavior, being instead fitted better by a power law MSD$(t) \sim t^{\alpha}$ with $\alpha \approx 0.22$ for $L=27$. Similarly, for $L=60$ we observe an algebraic growth of MSD with power $\alpha \approx 0.28$.

In conclusion, the numerical results presented in this section indicate that for $W=6.5$ and $U=12$, the $c$-boson spreads subdiffusively throughout the system, and MSD$(t) \sim t^{\alpha}$ at sufficiently large times and system sizes. This is in contrast with the localization of the $c$-boson reported, at \textit{the same values} of $W$ and $U$, in \cite{Brighi21b}. The latter results were inferred from numerical simulations for times up to $t_1=200$, which demonstrates the importance of a careful taking of the limit of large system sizes and long times when examining the slow dynamics of strongly disordered system. At the same time, we would like to emphasize that our results do not exclude the localization of the $c$-boson at \textit{larger values} of $W$  or $U$ at which the phenomenology discussed in \cite{Brighi21a, Brighi21b} could apply \footnote{We note that the region in which the localization in the considered system is perturbatively stable extends towards larger values of $W$ and $U$, as shown in \cite{Brighi21b}. However, increasing the disorder strength $W$ at fixed $U$ may lead to an increase of a pertubatively derived localization length of the $c$-boson, favouring \textit{delocalization} in the system. The value $W=6.5$ assumed in this work is close to the minimum of the localization length in a broad interval of $U$.}. Nevertheless, to obtain numerical results that can be unambiguously interpreted as indicating MBL or thermalization in the system for larger $W$ or $U$ requires even longer times and larger system sizes. This remains out of reach with our present computational resources. We note that the dynamics of the $c$-boson in the transient regime of the slow, logarithmic growth of MSD is analogous to dynamics of spin trapping effect relevant for impurities strongly interacting with one-dimensional clean systems \cite{Zvonarev07, Zvonarev09, Zvonarev09a, Imambekov08, Lamacraft09}. Moreover, a recent work \cite{Lezama22} finds a logarithmic growth of MSD that precedes approach to the final, infinite-temperature state in a one-dimensional Anderson insulator in the presence of a local noise.

  \begin{figure}
 \includegraphics[width=0.9\linewidth]{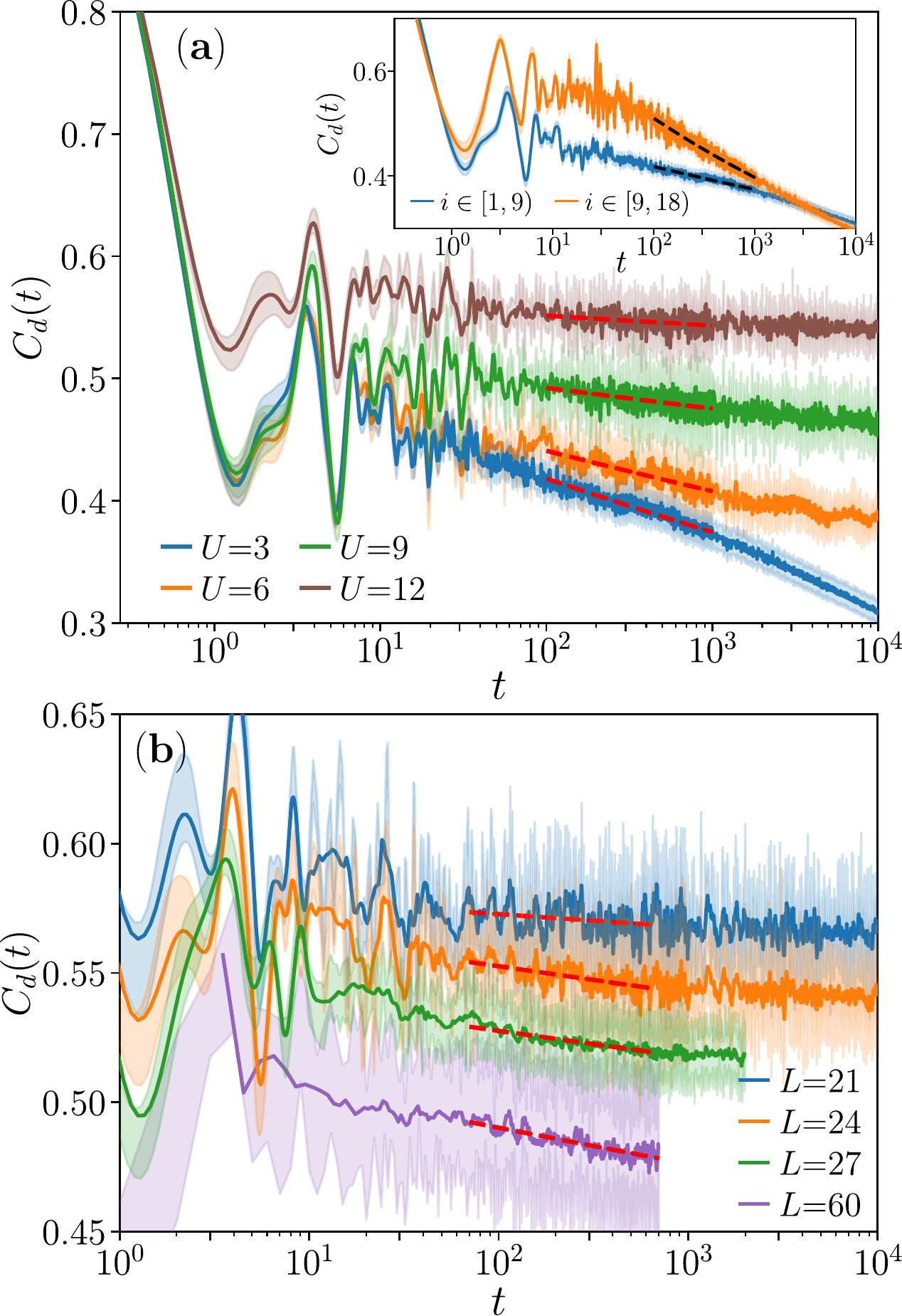}
  \vspace{-0.3cm}
  \caption{Slow dynamics of $d$-bosons, initially, the $c$-boson is placed on the leftmost site of the chain. (\textbf{a}): the density correlation function $C_d(t)$ for system size $L=24$ as function of time $t$ for various interaction amplitudes $U$ at fixed disorder strength $W=6.5$ for $l_1=1$ and $l_2=9$, the red dashed lines show the power-law fits $C_d(t) \sim t^{-\beta}$ with $\beta > 0$. The inset compares the correlation function for $U=3$ and $W=6.5$ in a left subsystem of the chain ($l_1=1$, $l_2=9$, blue line) and in the central part ($l_1=9$, $l_2=18$, orange line); (\textbf{b}): $C_d(t)$ for different system sizes $L$ at fixed $U=12$, $W=6.5$, and $l_1=1$, $l_2=9$. The dashed line shows power-law fits. For clarity, the results for $L=27$ and $L=60$ are shifted down respectively by $0.04$ and $0.07$. Data for $L=60$ was obtained with TDVP with bond dimension $\chi=800$.
  }\label{fig:CdLEF}
\end{figure}   

\section{Dynamics of the disordered bosons}
\label{sec:cor}
 
 In this section we investigate the dynamics of the $d$-bosons. In absence of interactions ($U=0$), the $d$-bosons form an Anderson insulator. For a non-vanishing $U$, the interactions with the $c$-boson induce a non-trivial many-body dynamics of $d$-bosons. To probe it, we consider the {disorder averaged} density correlation function {
 \begin{equation}
  C_d(t) = \frac{1}{D}\sum_{i=l_1}^{l_2}\braket{\psi(0) | \hat{n}_{i,d}(t) \hat{n}_{i,d} | \psi(0)},
 \end{equation} 
 where $\hat{n}_{i,d}(t) = e^{i H t} \ \hat{n}_{i,d} \ e^{-i H t}$ are the time-evolved number operators for the $d$-bosons in the Heisenberg representation, $\ket{ \psi(0) }$ is the initial state, taken either to be $\ket{\psi_{left}(0)}$ or $\ket{\psi_{cent}(0)}$), $D$ is a normalizing constant that assures $C_d(t=0)=1$, and $l_1, l_2$ limit the sites taken into account. For the initial product states considered in this work, the above formula simplifies into}
 \begin{equation}
  C_d(t) = \frac{1}{D}\sum_{i=l_1}^{l_2} n_{i,d}(t) n_{i,d}(0),
 \end{equation}
where $n_{i,d}(t)=\braket{\psi(t) | \hat{n}_{i,d}| \psi(t) }$, $\ket{\psi(t)} = e^{-i H t} \ket{\psi(0)}$ is the state of the system at time $t$. 
The density profile of the $c$-boson is highly non-uniform, hence, the various parts of the system thermalize at different rates. For that reason, we consider various choices of $l_1, l_2$ in the following. 

We start by fixing the system size to $L=24$ and take $\ket{\psi_{left}(0)}$ as the initial state of the system. We calculate the correlation function $C_d(t)$ taking $l_1=1$ and $l_2=9$, focusing on the part of the chain close to the initial position of the $c$-boson. The results are shown in Fig.~\ref{fig:CdLEF}(a). We observe a strong impact of the interaction strength $U$ on the decay of the correlation function and consequently on the thermalization of the system. The time evolution of the correlation function is well approximated in the interval $t\in[100,1000]$ by an algebraic decay, $C_d(t) \sim t^{-\beta}$ with $\beta \approx 0.048, 0.033, 0.015, 0.006$ respectively for $U=3,6,9,12$. For $U=3$, the decay of $C_d(t)$ becomes more rapid at longer times. For larger values of $U$, the slow power-law decay persists to longer times, similarly to what is observed for density correlation functions in disordered Heisenberg spin chain \cite{Sierant22}. Importantly, there is a strong relationship between the slow down of the spreading of the $c$-boson with increasing interaction strength $U$, recall Fig.~\ref{fig:clean_evo}(e), and the slow down of decay of density correlation function of $d$-bosons. Moreover, the dynamics of density correlations of $d$-bosons is highly non-uniform as shown in the inset of Fig.~\ref{fig:CdLEF}(a). Initially, the decay of $C_d(t)$ for $l_1=1$ and $l_2=9$ is faster than the decay in the middle of the system ($l_1=9$, $l_2=18$), but then, at times $t>100$, the decay in the middle becomes faster. A similar delay of the decay of $C_d(t)$ with the distance from the original position of the $c$-boson is also observed for larger values of $U$.

While the data for $L=24$ indicate the thermalization of the system for smaller values of $U$, the strongly interacting case, $U=12$, is less straightforward to understand: the power $\beta$ governing the decay of $C_d(t)$ is very small. To shed further light on the dynamics of the system for $U=12$ and $W=6.5$, we show $C_d(t)$ for various system sizes ranging from $L=21$ to $L=60$ in Fig.~\ref{fig:CdLEF}(b). We observe that the decay of $C_d(t)$ in the interval $t\in[100,1000]$ (or up to $t=700$ for $L=60$) is well approximated by a power-law with a coefficient $\beta$ that increases with system size: $\beta\approx 0.004, 0.008, 0.008, 0.011$ respectively for $L=21, 24, 27, 60$. This behavior of the $C_d(t)$ indicates a thermalization of the system at $U=12$ and $W=6.5$,  providing a complementary view  
on the dynamics of the system to the sub-diffusive growth of MSD of the $c$-boson reported in the preceding section. Importantly, the thermalization of the system slows down considerably with the distance from the original position of the $c$-boson. For instance, taking $l_1=10$, $l_1=18$ for $L=60$ we obtain, in the interval $t\in[100,700]$, that $C_d(t) \sim t^{-\beta}$ with exponent $\beta \approx 0.005$ that is approximately twice smaller than the exponent governing the decay of density correlation functions on the  $9$ leftmost sites of the chain. The thermalization of the system at $U=12$ and $W=6.5$ is an extremely slow process, even in the vicinity of the original position of the $c$-boson. Finally, we note that the results for the density correlation function $C_d(t)$ are fully analogous when $\ket{\psi_{cent}(0)}$ is used as the initial state of the system, the results are shown in App.~\ref{app:2}

\begin{figure}
 \includegraphics[width=0.9\linewidth]{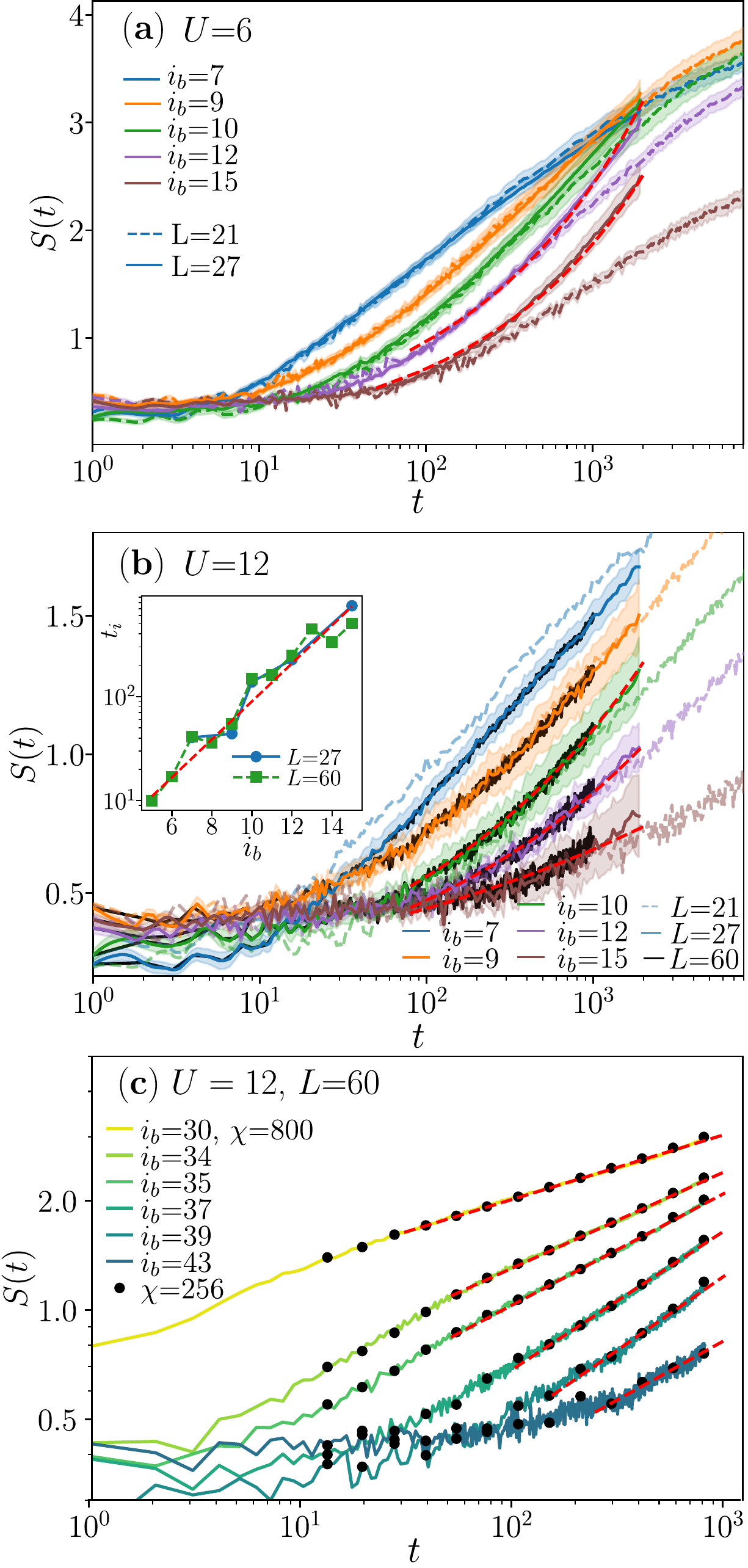}
  \vspace{-0.3cm}
  \caption{Entanglement entropy $S(t)$ in the system. (\textbf{a}): $S(t)$ for cut at bonds $i_b=7,9,10,12,15$, solid lines denote data for $L=27$, the dashed lines denote data for $L=21$, the red dashed lines show power-law fits $S(t)\sim t^{\gamma}$. The interaction strength $U=6$, disorder strength $W=6.5$, the initial state is $\ket{\psi_{left}(0)}$;
  (\textbf{b}): $S(t)$ for cuts at various $i_b$, colored solid lines denote data for $L=27$,  the dashed lines denote data for  for $L=21$, data for $L=60$ are denoted by black solid lines. The red dashed lines show power-law fits $S(t)\sim t^{\gamma}$, the interaction strength is $U=12$, disorder strength $W=6.5$, the initial state is $\ket{\psi_{left}(0)}$. The inset shows the time $t_i$ such that $S(t_i)=s_0$ versus the position of the cut $i_b$, the threshold value $s_0$ is taken as $0.6$, the dashed line shows an exponential fit $t_i \sim e^{0.42 i_d}$;
    (\textbf{c}): $S(t)$ for cuts at various $i_b$, colored solid lines denote data for $L=60$ and $\chi=800$ whereas black dots show data for $L=60$ and $\chi=256$, the initial state is $\ket{\psi_{cent}(0)}$. The red dashed lines show power-law fits $S(t)\sim t^{\gamma}$, note that log-log scale is used here.
  }\label{fig:entLEF}
\end{figure}

\section{Entanglement in the system}
\label{sec:ent}

In this section we investigate the entanglement entropy of the system. We consider a bipartition of the 1D lattice into two subsystems $A$ and $B$, so that $A\cup B$ is the full system and $A$ consists of sites $i=1,\ldots,i_b$. Using the time evolved state $\ket{\psi(t)}$, we calculate the reduced density matrix $\rho_A $ of the subsystem $A$ by tracing out the degrees of freedom associated with the subsystem $B$, $\rho_A = \mathrm{Tr}_{B}( \ket{\psi(t)}\bra{\psi(t)} )$ and obtain the von Neumann entanglement entropy as 
\begin{equation}
 S(t) = \mathrm{Tr} \left[ \rho_A \ln \rho_A \right].
\end{equation}
To simplify the numerical calculations, we employ the conservation of the total number of $c$-bosons and $d$-bosons  but we do not resolve the associated number entropies \cite{Schuch04,Schuch04b,Donnelly12,Turkeshi20,Lukin19,Sierant19c}. 

To start our investigation of dynamics of entanglement entropy in the regime of slow dynamics in the considered system, we set $U=6$ and calculate $S(t)$ for system sizes $L=21, 27$ taking $\ket{\psi_{left}(0)}$ as the initial state. The results, shown in Fig.~\ref{fig:entLEF}(a), indicate a power-law growth of entanglement entropy $S(t) \sim t^{\gamma}$, similarly to the regime of slow dynamics close to the MBL regime \cite{Luitz16}. Close to the original position of the $c$-boson, for bonds $i_b \leq 10$, the entanglement entropy starts to saturate in time to a value of the order of the thermal value, consistently with the thermalization of the system in the limit of long times and large $L$. The time at which $S(t)$ starts to grow increases with the distance from the leftmost site of the chain. In particular, for $i_b=15$ and $L=21$, the entanglement entropy $S(t)$ seems to follow a logarithmic in time growth, similarly as in the MBL regime of disordered Heisenberg spin chain \cite{Znidaric08, Bardarson12}. However, when the system size is increased to $L=27$, we observe a power-law growth in the interval of times $t\in[50,2000]$, with power $\gamma \approx 0.42$. Similarly, for $i_b=12$ we get $\gamma \approx 0.40$. Therefore, for $U=6$, we see clear signatures of thermalization in the system. 

The behavior of entanglement entropy is similar also for $U=12$. While the results for $i_b=7,9$ could suggests a logarithmic increase of $S(t)$, the growth of $S(t)$ is more rapid for $i_b \geq 10$. Performing fits in the interval $t\in[80,2000]$ we find a power-law dependence $S(t) \sim t^{\gamma}$ with $\gamma \approx 0.29, 0.26, 0.17$ respectively for $i_b = 10,12,15$. Interestingly, in the interval of times for which data for $L=60$ are available, we find that $S(t)$ for $L=27$ and $L=60$ almost overlap, see the black lines in Fig.~\ref{fig:entLEF}(b). Finally, we note that the value of $S(t) \approx 1.5$ at $t=1000$ (for $i_b=7$) is comparable to the value ($S\approx 1.7$) of the entanglement entropy obtained after $t=1000$ tunneling times in the disordered Heisenberg spin chain at $W_{XXZ}=4$ \cite{Chanda20t}. At that value of the disorder strength the disordered Heisenberg spin chain is believed to be still in the thermal phase \cite{Sierant20p, Morningstar22}. Notably, this value of $S(t)$ is significantly larger than the value  $S(t)\approx 0.25$ of entanglement entropy observed at similar times in disorder Heisenberg chain at disorder strength $W_{XXZ}=10$ at which one still observes a non-trivial dynamics of density correlation functions at experimentally relevant time scales \cite{Sierant22}.

In conclusion, the time evolution of the entanglement entropy $S(t)$ offers a third perspective,  along the behavior of MSD$(t)$ and $C_d(t)$, on the thermalization that occurs in the system at $U \leq 12$. The dependence of the moment of the start of entanglement increase on the distance between the bond and the initial position of the $c$-boson reflects the high non-uniformity of the initial condition. This rapidly introduces long-time scales into the dynamics of the system. For instance, the time at which the entanglement entropy $S(t)$ exceeds a threshold value $s_0=0.6$ increases exponentially with the distance from the left side of the chain, as shown in the inset of Fig.~\ref{fig:entLEF}(b). In particular, already for $i_b=22$, up to the time $t=1000$ reached for $L=60$, we observe 
no traces of the increase of $S(t)$ beyond the value characteristic for Anderson insulator (admitted after $t\sim 5$). This highlights that the process of thermalization of the system is very slow in the considered parameter regime, and parts of the system in which the density of the $c$-boson is negligibly small remain in Anderson insulating state at practically relevant timescales. Finally, we note that we observe a similar behavior of the entanglement entropy when $\ket{\psi_{cent}(0)}$ is used as the initial state. The entanglement entropies for $L=60$ obtained TDVP algorithm, shown in Fig.~\ref{fig:entLEF}(c), practically overlap for bond dimensions $\chi=256$ and $\chi=800$, confirming the convergence of the algorithm with $\chi$. The data are well approximated by a power-law dependence  $S(t)\sim t^{\gamma}$ in time interval $t\in[t_{min},800]$, respectively with $t_{min}=30$, $\gamma \approx 0.18$ for $i_b=30$; $t_{min}\approx 50$, $\gamma=0.27$ for $i_b=34$; $t_{min}=50$, $\gamma \approx 0.31$ for $i_b=35$;
 $t_{min}=100$, $\gamma \approx 0.38$ for $i_b=37$;
  $t_{min}=150$, $\gamma \approx 0.40$ for $i_b=39$;
   $t_{min}=240$, $\gamma \approx 0.33$ for $i_b=43$.
The power-law fits to $S(t)$, consistent with the thermalization of the system work in wider time intervals than logarithmic fits (which are consistent with MBL in the system). Reaching times beyond $t=200$ is crucial to  demonstrate this behavior. For instance, looking at data for $i_b=34$, we see a downward curvature in the interval $t\in[20,200]$. However, the curvature disappears as one considers the $S(t)$ in the wider time interval  $t\in[50,800]$. This highlights the importance of taking the long time limit when assessing the properties of slow dynamics in strongly disordered systems and
is the main source of discrepancies between the results of the present work and of the works \cite{Brighi21a, Brighi21b} at the parameters $U=12$ and $W=6.5$.

\begin{figure}
 \includegraphics[width=0.9\linewidth]{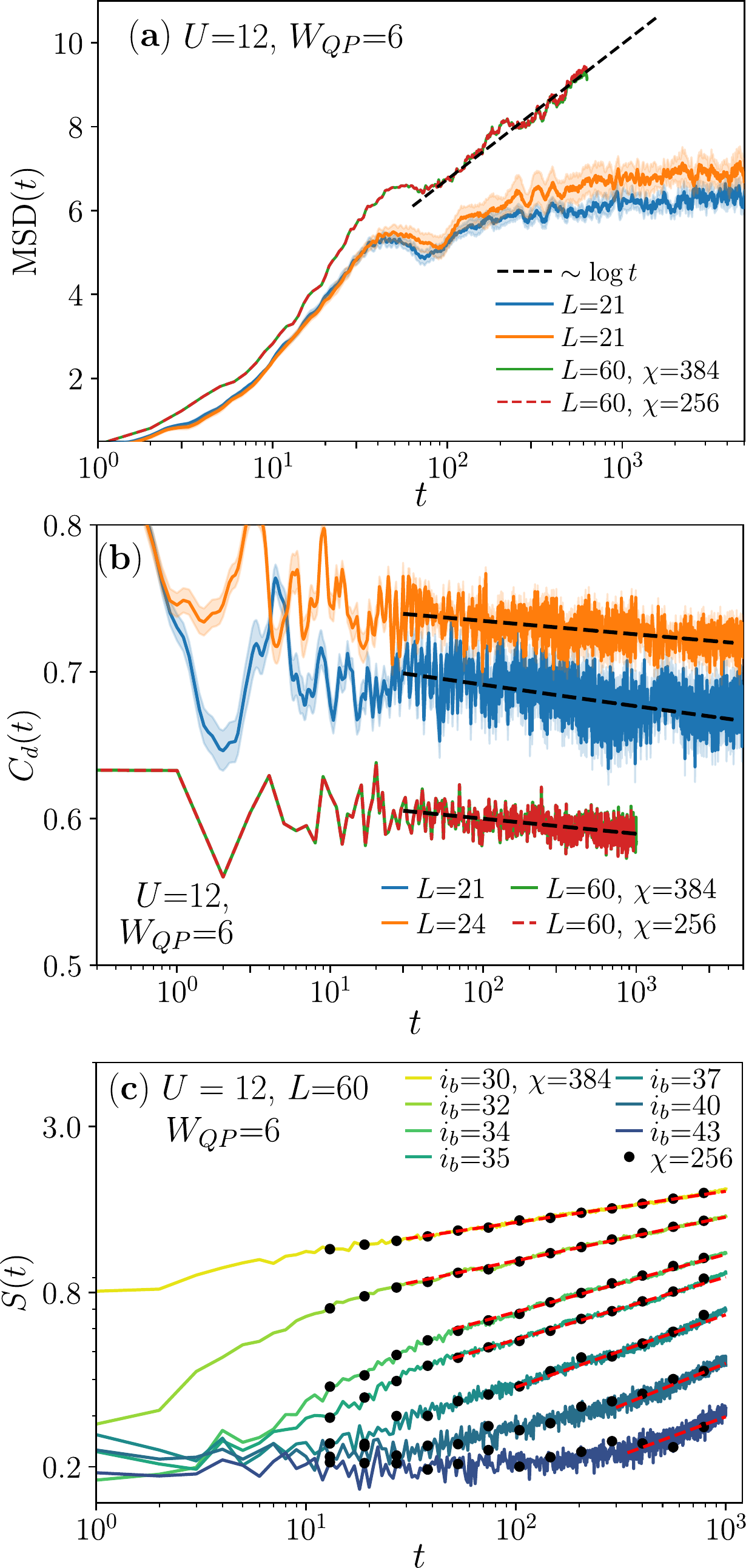}
  \vspace{-0.3cm}
  \caption{Slow dynamics in presence of quasi-periodic potential with amplitude $W_{QP}=6$  for initial state $\ket{\psi_{left}(0)}$ and interaction strength $U=12$.
  (\textbf{a}): The MSD$(t)$ as function of time for various system sizes (for TDVP data, at $L=60$, the bond dimension $\chi$ is indicated). While for $ L \leq 24$ we observe a saturation of MSD, the data for $L=60$ are well approximated by MSD$(t) \sim \log(t) $;  (\textbf{b}): the density correlation function $C_d(t)$ calculated around the middle site of the chain (for $i\in[L/2-3, L/2+3]$) for various system sizes, fitted with a power-law decay $C_d(t) \sim t^{-\beta}$. (\textbf{c}): The entanglement entropy $S(t)$ for various cuts $i_b$ and system size $L=60$. Data obtained with TDVP with bond-dimension $\chi=384$ are denoted by solid lines, whereas data obtained with $\chi=256$  are denoted by dots. The dashed lines show power-law fits $S(t) \sim t^{\gamma}$ with $\gamma \approx 0.11, 0.15, 0.20, 0.22, 0.25, 0.29, 0.27$ respectively for $i_b=30, 32, 34, 35, 37, 40, 43$.
  }\label{fig:qpd}
\end{figure}   

\section{Slow dynamics in a quasi-periodic potential}
\label{sec:qpd}

We now turn to investigation of the dynamics of the system in presence of quasiperiodic potential 
$h_j = W_{QP} \cos(2\pi k j + \phi)$ described in Sec.~\ref{sec:model}. To consider the regime of slow dynamics we fix the interaction strength as $U=12$, and we set $W_{QP}=6$.

We start by studying the spreading of the $c$-boson. The MSD$(t)$, shown in Fig.~\ref{fig:qpd}(a) nearly saturates at system size independent value $\approx 7$ for $L=21$ and $L=24$. For that reason one could infer that the $c$-boson remains localized in the quasiperiodic potential. However, this is not the case in the large system size limit as the data for $L=60$ shows. Instead, we observe a slow increase of MSD in time, which is well fitted with a logarithmic time dependence. The overlapping of data for $\chi=256$ and $\chi=384$ indicates that the results of TDVP algorithm are well converged up to time $\approx 500$. A comparison with results for the disordered case, together with subdiffusive behavior of of MSD$(t)$ at smaller values of $U$, suggests that at even larger times and system sizes the logarithmic increase of MSD$(t)$ will be replaced by a power-law dependence at $U=12$.

The behavior of the density correlation function $C_d(t)$ around the initial position of the $c$-boson ($l_1=L/2-3$, $l_2=L/2+3$), presented in Fig.~\ref{fig:qpd}(b),  provides complementary perspective on the dynamics of the system. We observe a slow decay of $C_d(t)$, well fitted by a power-law $C_d(t) \sim t^{-\beta}$
with $\beta \approx 0.006$, $\beta \approx 0.004$, $\beta \approx 0.008$ respectively for $L=21, 24, 60$ and time intervals $t\in[30,t_{max}]$ with $t_{max}=4000, 4000, 1000$ (note that the data for $C_d(t)$ are converged to larger times than MSD$(t)$). Finally, similarly to the disordered case, the entanglement entropy $S(t)$ increases according to a power-law in time, $S(t)\sim t^{\gamma}$, as shown in Fig.~\ref{fig:qpd}(c). 

All in all, the results for quasiperiodic potential indicate a presence of a regime of slow dynamics similar to the case of disordered system. The gradual thermalization of the system is manifested by subdiffusive spreading of the $c$-boson, the power-law decay of density correlation function of $d$-bosons and a power-law increase of the entanglement entropy.

\section{Conclusions}
\label{sec:con}

We have considered a single mobile impurity interacting with an Anderson insulator and investigated a regime of slow dynamics that appears in the system in presence of strong disorder
and interactions. To probe the dynamics of the system, we have studied time evolution of MSD of the impurity, the density correlation functions of the particles subject to disorder that initially form the Anderson insulator as well as the entanglement entropy. 
Due to the slowness of the dynamics of the system, we find that reaching times beyond $t=200J^{-1}$ is crucial for understanding the physics of the system for interaction strength $U=12$ and disorder amplitude $W=6.5$, considered previously in \cite{Brighi21a, Brighi21b}. Combining numerically exact simulations at system sizes $L\leq30$ with tensor network approaches at larger $L$, we observe thermalization of the system at longer times for that parameters. In particular, we find that MSD grows algebraically in time, consistently with sub-diffusive spreading of the impurity. This behavior is intercontected with a slow, consistent with a power-law in the considered time interval, decay in time of the correlation functions of the particles subject to disorder. Consequently, the entanglement entropy grows algebraically in time. We have also considered smaller values of interaction strength $U$ showing that the regime of slow dynamics extends in a wide parameter space of the system. Finally, we have demonstrated an appearance of a similar regime of slow dynamics in the quasiperiodic potential.

The slow dynamics of the impurity interacting with Anderson insulator is reminiscent of the regime of the slow dynamics preceding the MBL phase \cite{Luitz16}. Therefore, similar problems appear in both cases when one tries to understand the behavior of the system in the asymptotic limit of large times and system sizes \cite{Sierant22}. Our simulations reveal the tendencies towards thermalization of the system (at $U=12$ and $W=6.5$) on experimentally relevant time scale $t=1000J^{-1}$. The trends in the behavior of MSD, correlation functions and entanglement entropy suggest that the dynamics of the system speeds up when time and length scales are increased (for instance the logarithmic growth of MSD crossover for system size $L=27$ crossovers to a power-law growth at $L=60$). On the other hand, our results do not exclude a scenario in which a certain increase of $U$ or $W$  will lead to even bigger slow down of the dynamics of the system that will lead to an asymptotic localization of the impurity and MBL of the disordered particles, as discussed in \cite{Brighi21a, Brighi21b}.

Regardless of the asymptotic fate of the system, the slow spreading of the impurity in the considered dynamical regime induces thermalization of the Anderson insulator at time scale that increases exponentially with the distance from the initial position of the impurity. This indicates that parts of the Anderson insulator distant from the initial position of the impurity will remain in the insulating state at practically relevant time scales. Finally, we would like to note that the considered system is interesting in the context of the avalanche mechanism of destabilization of MBL \cite{DeRoeck17, Luitz17}. From that perspective, the impurity is a $L$-level system that induces a thermalizing avalanche in the system. Our result emphasize the reach physics of that process that appears on relatively long but experimentally relevant time scales.

\acknowledgements    

We acknowledge the support of  PL-Grid Infrastructure. The work of J.Z. was funded by  National Science Centre (Poland) within  Opus grants 2019/35/B/ST2/00034 and  2021/43/I/ST3/01142.  P.S. and M.L acknowledge support from: ERC AdG NOQIA; Ministerio de Ciencia y Innovation Agencia Estatal de Investigaciones (PGC2018-097027-B-I00/10.13039/501100011033, CEX2019-000910-S/10.13039/501100011033, Plan National FIDEUA PID2019-106901GB-I00, FPI, QUANTERA MAQS PCI2019-111828-2, QUANTERA DYNAMITE PCI2022-132919, Proyectos de I+D+I “Retos Colaboración” QUSPIN RTC2019-007196-7); MCIN Recovery, Transformation and Resilience Plan with funding from European Union NextGenerationEU (PRTR C17.I1); Fundació Cellex; Fundació Mir-Puig; Generalitat de Catalunya (European Social Fund FEDER and CERCA program (AGAUR Grant No. 2017 SGR 134, QuantumCAT \ U16-011424, co-funded by ERDF Operational Program of Catalonia 2014-2020); Barcelona Supercomputing Center MareNostrum (FI-2022-1-0042); EU Horizon 2020 FET-OPEN OPTOlogic (Grant No 899794); EU Horizon Europe Program (Grant Agreement 101080086 — NeQST), National Science Centre, Poland (Symfonia Grant No. 2016/20/W/ST4/00314); European Union’s Horizon 2020 research and innovation program under the Marie-Skłodowska-Curie grant agreement No 101029393 (STREDCH) and No 847648 (“La Caixa” Junior Leaders fellowships ID100010434: LCF/BQ/PI19/11690013, LCF/BQ/PI20/11760031, LCF/BQ/PR20/11770012, LCF/BQ/PR21/11840013). Views and opinions expressed in this work are, however, those of the authors only and do not necessarily reflect those of the European Union, European Climate, Infrastructure and Environment Executive Agency (CINEA), nor any other granting authority. Neither the European Union nor any granting authority can be held responsible for them.  For the purpose of Open Access, J.Z. has
applied a CC-BY public copyright licence to any AuthorAccepted Manuscript (AAM) version arising from this submission.

\appendix

\section{Convergence of numerical results}
\label{app:1}

In this section we discuss the convergence of our numerical results. The results for $L\leq 30$, obtained with expansion of the evolution operator $e^{-i H t}$ into Chebyshev polynomials are numerically exact, the details of the scheme are given in \cite{Sierant22}. On the other hand, the tensor network based TDVP algorithm approximates the time evolved state $\ket{\psi(t)}$ with accuracy that depends on the bond dimension $\chi$ of the underlying MPS state \cite{Schollwoeck11}. In order to probe the convergence of the algorithm, 
we have shown results for two bond dimension in Fig.~\ref{fig:entLEF} and Fig.~\ref{fig:qpd}. The data for entanglement entropies (Fig.~\ref{fig:entLEF}(c) for $\chi=256,800$) as well as for MSD$(t)$, $C_d(t)$ and $S(t)$ (Fig.~\ref{fig:qpd} for $\chi=256,384$) practically overlap, indicating that the numerical results are well converged with the bond dimension $\chi$. 

\begin{figure}[ht]
 \includegraphics[width=0.92\linewidth]{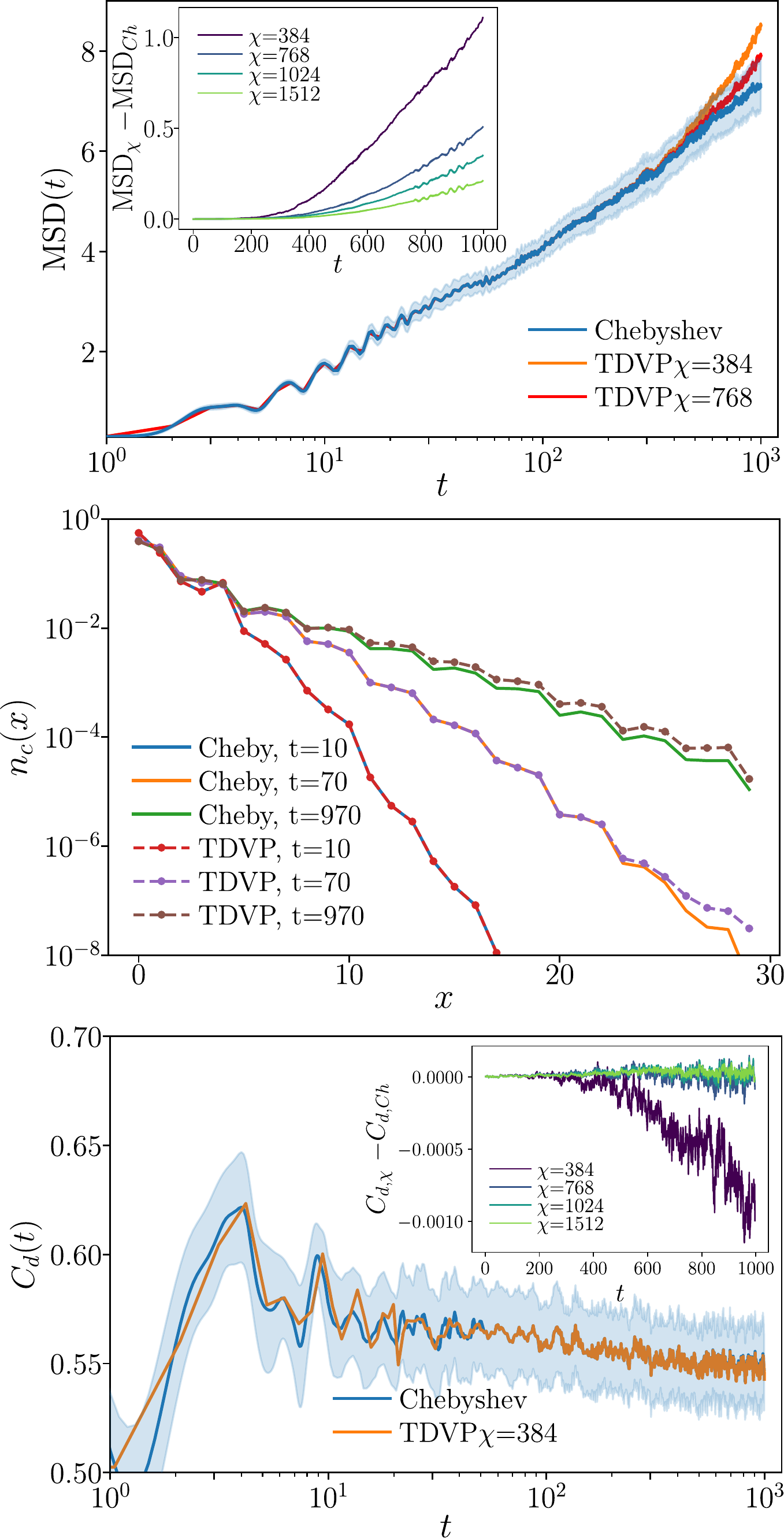}
  \vspace{-0.3cm}
  \caption{Convergence of TDVP vs numerically exact results from Chebyshev time propagation scheme. System size $L=30$, interaction strength $U=12$, disorder amplitude $W=6.5$, initially, the c-boson ocucpies the left-most site of the chain. Top: MSD as function of $t$ obtained with Chebyshev method (statistical error, calculated by resampling over disorder realizations is denoted by shaded blue region) is compared against TDVP results. The inset shows the difference between TDVP result ($MSD_{\chi}$) and Chebyshev data ($MSD_{CH}$); Middle: density of the clean boson $n_c(x)$ as function of the site number $x$ for various times $t=10,70,970$, Chebyshev results denoted by solid lines, TDVP results by dashed dotted lines; Bottom: the same as top panel, but for the density correlation function $C_d(t)$ of d-bosons. 
  }\label{fig:convergenceL30}
\end{figure}

In order to further illustrate the convergence of TDVP, we compare numerically exact results obtained with Chebyshev time propagation for $L=30$ with TDVP results. As we show in Fig.~\ref{fig:convergenceL30}(a), the MSD$(t)$ is predicted with accuracy $1\%$ up to time $t\approx 400$ for $\chi=384$ and $t\approx 700$ for $\chi=768$. The errors in MSD$(t)$ are associated with overestimation of the tails of the density profile $n_c(x)$ of the $c$-boson as shown if Fig.~\ref{fig:convergenceL30}(b). Nevertheless, up to times $t\approx 800$, the quantitative, logarithmic in time, behavior of MSD$(t)$ is well reproduced by TDVP algorithm. The same applies to data presented in Fig.~\ref{fig:clean_evo}. Finally, Fig.~\ref{fig:convergenceL30}(c) shows that the convergence of the density correlation function $C_d(t)$ with $\chi$ is much better and already $\chi=384$ practically reproduces the exact result.

\section{Dynamics of disordered bosons - further results}
\label{app:2} 

\begin{figure}[ht]
 \includegraphics[width=0.9\linewidth]{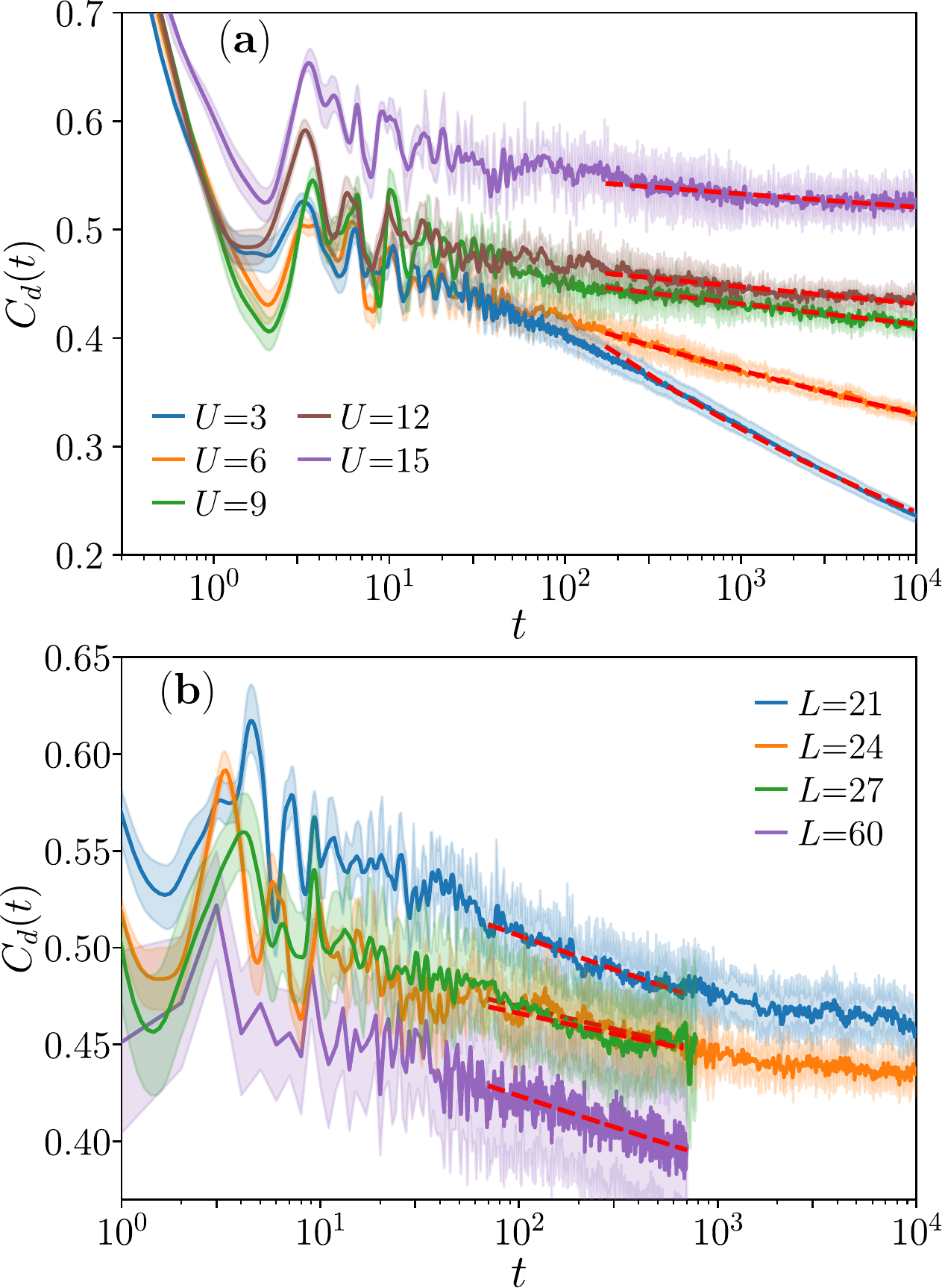}
  \vspace{-0.3cm}
  \caption{Slow dynamics of $d$-bosons, initially, the $c$-boson is placed in the middle of the chain. (\textbf{a}): the density correlation function $C_d(t)$for system size $L=24$ as function of time $t$ for various interaction amplitudes $U$ at fixed disorder strength $W=6.5$, the red dashed lines show the power-law fits $C_d(t) \sim t^{-\beta}$ with $\beta > 0$; (\textbf{b}): $C_d(t)$for different system sizes $L$ at fixed $U=12$, $W=6.5$. The dashed line shows power-law fits. For clarity, the results for $L=60$ are shifted down by $0.07$.
  }\label{fig:CdCEN}
\end{figure}

In this section we consider the density correlation function $C_d(t)$ of the $d$-bosons, calculating it close to the center of the chain: $l_1 = L/2-3$, $l_2 = L/2+3$. We find that $C_d(t)$ decays in time, and that the decay is well approximated by a power-law: $C_d(t)\sim t^{-\beta}$. For fixed system size $L=24$, see \ref{fig:CdCEN}(a), the decay slows down considerably with $U$: we obtain $\beta \approx 0.12, 0.05, 0.022, 0.015, 0.010$ respectively for $U=3,6,9,12$. Fixing the interaction strength as $U=12$, we find $\beta \approx 0.031, 0.022, 0.021, 0.030$ respectively for $L=21, 24, 27, 60$ (performing the fit in interval $t\in[100,700]$). The latter suggests that the decay of $C_d(t)$ is converged with system size at the considered time scales. It also confirms the gradual thermalization of the system at $U=12$ and $W=6.5$.

\normalem

%

\end{document}